\documentclass[journal]{IEEEtran}
\usepackage{amsmath,amsfonts}
\usepackage{algorithmic}
\usepackage{algorithm}
\usepackage{array}
\usepackage[caption=false,font=normalsize,labelfont=sf,textfont=sf]{subfig}
\usepackage{textcomp}
\usepackage{stfloats}
\usepackage{url}
\usepackage{verbatim}
\usepackage{graphicx}
\usepackage{cite}
\usepackage{xcolor}
\begin{document}

%\title{When semantic Communication Meets Large Language Models (LLMs) in Edge Computing-Based IoT Systems}
%\title{Next-Gen IoT: semantic Communication Using Large Language Models (LLMs) at the Edge of IoT Networks}
\title{Large Language Models (LLMs) for Semantic Communication in Edge-based IoT Networks}

\author{Alakesh Kalita,~\IEEEmembership{Senior Member,~IEEE,}
        % <-this % stops a space
\thanks{}% <-this % stops a space
\thanks{}
\thanks{}
\thanks{}
}



\maketitle

\begin{abstract}
With the advent of Fifth Generation (5G) and Sixth Generation (6G) communication technologies, as well as the Internet of Things (IoT), semantic communication is gaining attention among researchers as current communication technologies are approaching Shannon's limit. On the other hand, Large Language Models (LLMs) can understand and generate human-like text, based on extensive training on diverse datasets with billions of parameters. Considering the recent near-source computational technologies like Edge, in this article, we give an overview of a framework along with its modules, where LLMs can be used under the umbrella of semantic communication at the network edge for efficient communication in IoT networks. Finally, we discuss a few applications and analyze the challenges and opportunities to develop such systems.
\end{abstract}


\section{Introduction}
\begin{figure*}[!t]
\centering
\includegraphics[width=.8\textwidth]{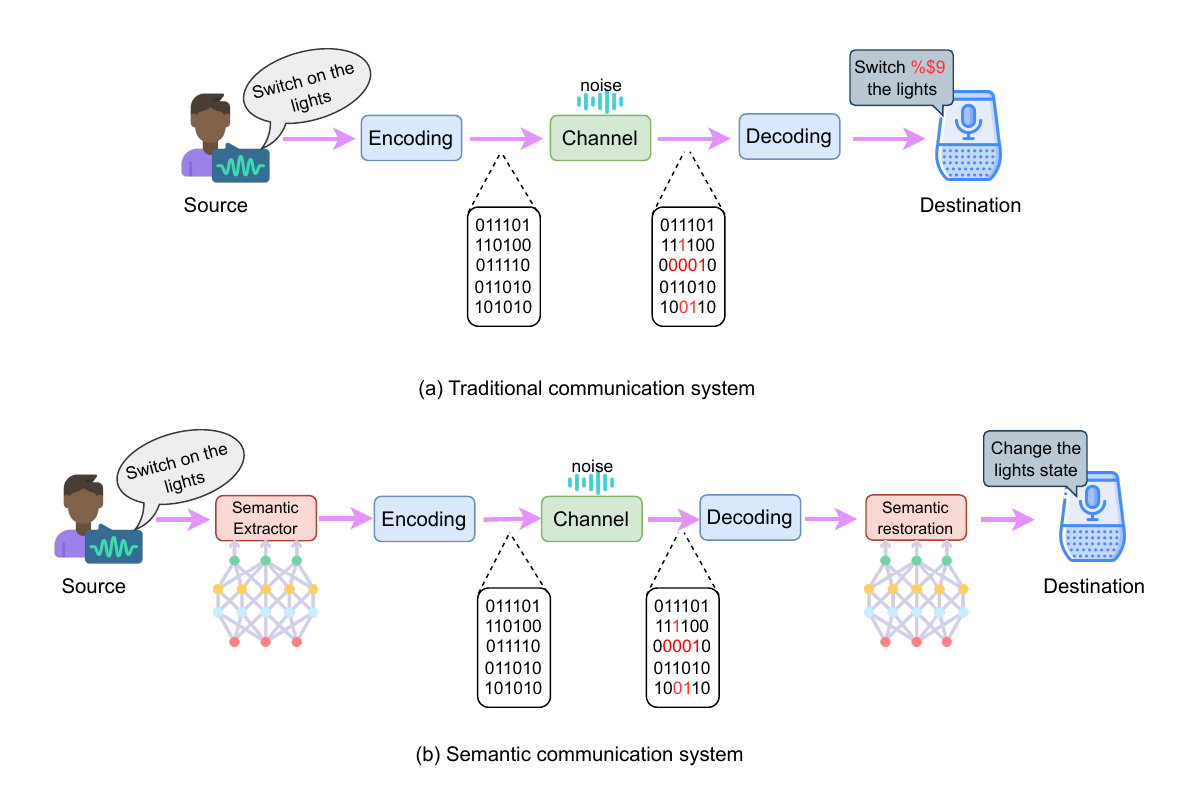}
\caption{Working of (a) Transitional communication and (b) semantic communication. The latter ensures bit-wise (syntax) transmission, whereas semantic communication ensures the meaning of the transmitted information and works even if there are some bit-level errors in the receiver.}
\label{fig_1}
\end{figure*}

\IEEEPARstart{T}{he} traditional communication system is mainly concerned with the accurate transmission of the bits (by transforming into electromagnetic signal) from one host to another host. Even though using modern encoding, decoding, modulation, demodulation, etc., techniques bits are correctly sent to the other end, due to the increasing number of Internet-connected devices in the world, data generated by AR/VR devices, \textit{Internet of Things} (IoT) devices, Shannon's capacity is already about to reach \cite{shannon1949mathematical}. Therefore, nowadays researchers are exploring semantic communications, which refers to the process of exchanging information where the focus is on the meaning and interpretation of the transmitted content, rather than just the accurate (i.e., any bit-level error) transmission of raw data \cite{9852388, 9955525}. In brief, traditional communication focuses on syntax and the physical transmission of data. In contrast, semantic communication ensures the exchanging of the inherent meaning of the information \cite{9679803, grassucci2024generativeaimeetssemantic}. In Figure~\ref{fig_1}, we show how traditional communication is different from semantic communication, where a person asks his smart device (e.g., Alexa, Google Nest, Siri etc.,) to switch on his lights in a noisy environment. In traditional communication, if the transmitted bits get corrupted due to external noise, the receiver or the destination cannot understand the user command. Hence, no action will be taken. However, in semantic communication, the destination can try to understand/infer the user command even if some of the bits get corrupted along the way and based on retrieved meaning, action will be taken. 

To transform traditional communication into semantic communication, recent Artificial Intelligence (AI) technologies such as \textit{Deep Neural Networks} (DNN), and \textit{Recurrent Neural Networks} (RNN)  show exciting results in extracting and restoring the semantic meaning of the transmitted information \cite{9398576}. However, deep generative-based \textit{Large Language Models} (LLMs) are designed to understand and generate human language. Recent LLMs, such as GPT-4, GPT-4o, and Llama 3.1  are trained on vast amounts of text data, enabling them to perform a wide range of language tasks, including translation, summarization, and conversation. Recently, OpenAI's GPT series, including models like ChatGPT, has garnered significant attention due to the advancements in LLMs which can generate coherent and contextually relevant responses to users' queries, making them valuable tools in various applications like customer service, content creation, and research. Therefore, LLMs can be a great tool for semantic communications for both feature extractions and restorations \cite{10328186}.

With the development of digitization, IoT technology has become a prominent topic in the field of networking and communication. IoT aims to connect billions of physical devices which can be both living and non-living entities found on the earth's surface with sensing/actuation and communication capacity. IoT technology enhances automation, improves efficiency, and provides real-time insights across various sectors, including healthcare, agriculture, transportation, and smart homes. By enabling seamless connectivity and data exchange, IoT is transforming how we interact with the world around us \cite{10.1145/3536166}. Now, the integration of semantic communication with LLMs in IoT networks can make a huge transformation towards more intelligent, efficient, and responsive IoT systems \cite{lin2024pushinglargelanguagemodels}. This convergence leverages the natural language understanding capabilities of LLMs to enhance the semantic communication framework in IoT, allowing devices to communicate in a more meaningful and context-aware manner. For example, if a user says ``\texttt{Set up for movie night}'', semantic communication helps to \textit{set the Dim lights in the living room}, \textit{close the blinds, adjust the thermostat to a comfortable temperature, turn on the TV and navigate to the user's preferred streaming service}. All can be done with a single command, which requires multiple individual commands for each actuation in the current IoT systems. Thus, LLM-based communication in IoT can significantly improve efficiency, bandwidth utilization, and user interactions. 

However, the devices used in IoT networks are resource-constrained in terms of processing capacity, memory, and power. Therefore, using LLM in this kind of device is not practically trivial even at the border router (or sometimes in IoT gateways) of the IoT networks. Similarly, technologies like cloud computing have their disadvantages like higher latency and cost. Therefore, edge computing can be a suitable approach to use LLM under the umbrella of semantic communication for IoT applications such as smart homes, smart industries, etc. Edge computing processes data closer to the source of the data generators rather than relying on centralized data centres at higher propagation delay \cite{10342693}. This approach reduces latency, conserves bandwidth, and enhances real-time data processing capabilities. Hence, in the context of IoT, edge computing plays a crucial role by enabling devices to analyze and respond to data locally.

Therefore, in this article, we briefly discuss how LLMs can be used as a part of semantic communication for establishing efficient communication in edge computing-based IoT systems. In the next section, we provide basic background information on semantic communication, LLMs, IoT, and edge computing. After that, we discuss our proposed framework where LLMs can be used for semantic communication in Edge-based IoT systems. Finally, before concluding our work, we briefly discuss some of its applications, and challenges and opportunities in such systems.

\section{Background}
\subsection{Semantic Communication}
Traditional communication systems focus on providing high data transmission rates and minimizing symbol (bit) error rates. In such system, Shannon only focused on the technical level, which deals with encoding the source message into a bit sequence and recovering the same sequence at the receiver passing through a noisy channel without considering the underlying meaning of the message. In contrast, recently, a new paradigm has emerged in wireless communications, which shifted the focus of communication towards the semantic level, leading to the development of semantic communication. Semantic communication targets the exchanging of semantically equivalent information without necessarily requiring an identical match to the original transmitted information.  The core concept of semantic communications is to extract the ``\texttt{meanings}'' or ``\texttt{features}'' from the transmitted information at the \textit{source} and to ``\texttt{interpret}'' this semantic information at the \textit{destination}. However, semantic communication requires conventional methods to encode relevant information. To extract the meanings or features, semantic communication should have a \textit{Knowledge Base} (KB), which needs to be updated periodically based on the information it receives from the user's input to retrieve and interpret new information as user inputs keep on changing. Overall, the semantic communication system is complex, where both the source and destination need to use the KB using advanced DNN or RNN techniques along with the functions of traditional communication. Additionally, the semantic source and destination can perceive their environment and operate autonomously.

\subsection{Large Language Models (LLMs)}
LLMs are advanced AI-based systems designed to understand and generate human-like text based on vast amounts of data. The training process of these kinds of AI models requires adjusting millions or even billions of parameters through a technique called \textit{back-propagation}, and finally, optimizes the models' ability to understand and generate language. During training, LLMs are exposed to large datasets containing diverse text sources, which allows them to learn the statistical properties of language. After training, the model can generate text by taking an initial prompt as input and predicting subsequent words using diffusion techniques. This approach allows the model to iteratively refine its predictions, maintaining coherence and context throughout the generated sequence based on its learned parameters. Fine-tuning specific tasks or domains can further enhance the model's performance, making it adept at specialized applications such as technical support, content generation, or conversational agents. The ability to handle diverse linguistic patterns and contexts makes LLMs a powerful tool for a wide range of natural language processing tasks. Recent LLM architectures like GPT-4 and Llama 3.1, can generate coherent and contextually relevant responses, making them valuable tools for applications in natural language processing and human-computer interaction.

\subsection{Edge Computing for LLMs}

Edge computing is a distributed computing paradigm that brings computation and data storage closer to the location where it is needed and where the data is generated. Thus, the latency for processing sensory data and receiving a response has significantly reduced compared to other distributed computing approaches, such as cloud computing. In brief, edge computing improves response times and also saves bandwidth by processing the data locally and not sending the requests to the cloud. Thus, Edge computing can significantly enhance the performance and utility of LLMs by bringing computation and data storage closer to the data sources, thereby reducing latency and bandwidth usage. By deploying LLMs on edge devices, it is possible to process data locally rather than relying on centralized cloud servers. Additionally, edge computing improves privacy and security by keeping sensitive data on local devices rather than transmitting it over the Internet.  Moreover, it enables better scalability, as processing loads are distributed across multiple edge nodes rather than concentrated in a central cloud, which can prevent bottlenecks and reduce the risk of downtime of the central cloud server.

\subsection{Internet of Things (IoT)}
IoT refers to a network of interconnected physical devices that communicate and exchange data with each other and distributed computing systems such as Edge and Cloud over the Internet. Apart from having sensors (to sense the environment) and actuators (to act on the environment), IoT devices must have a communication facility for connectivity. However, due to the resource-constrained nature of IoT devices, heavy processing can not be done within them. Therefore, IoT networks mainly collect, send, and receive data for a wide range of applications to provide insights on data, automate processes, and improve efficiency. The integration of IoT with edge computing further enhances its capabilities by allowing data to be processed locally on nearby edge devices, leading to faster decision-making and reduced dependency on centralized cloud servers. The integration of IoT and edge computing is crucial for applications requiring low latency, high reliability, and real-time data processing.

Thus, semantic communication, LLMs, edge computing, and IoT networking can work together to create intelligent, user-friendly, efficient, and quick responsive systems. In such a system, IoT devices collect data from the environment and transmit it to edge computing nodes for immediate processing. After that, offline LLM in the edge device can analyze and interpret the received data as a part of semantic communication to understand the context and intent of the transmitted data. Hence, such an architecture can enable real-time, context-aware responses and decisions, a user-friendly and efficient IoT network eco-system to improve the already available IoT applications further. %\textcolor{red}{For example, in a smart city, sensors (IoT devices) can monitor traffic conditions, edge computing nodes process this data locally, and LLMs interpret the information to provide real-time traffic management solutions and alerts.}

\section{A Framework for LLM-Based semantic Communication in IoT Networks}
\subsection{System Model}
\begin{figure*}[!t]
\centering
\includegraphics[width=.9\textwidth]{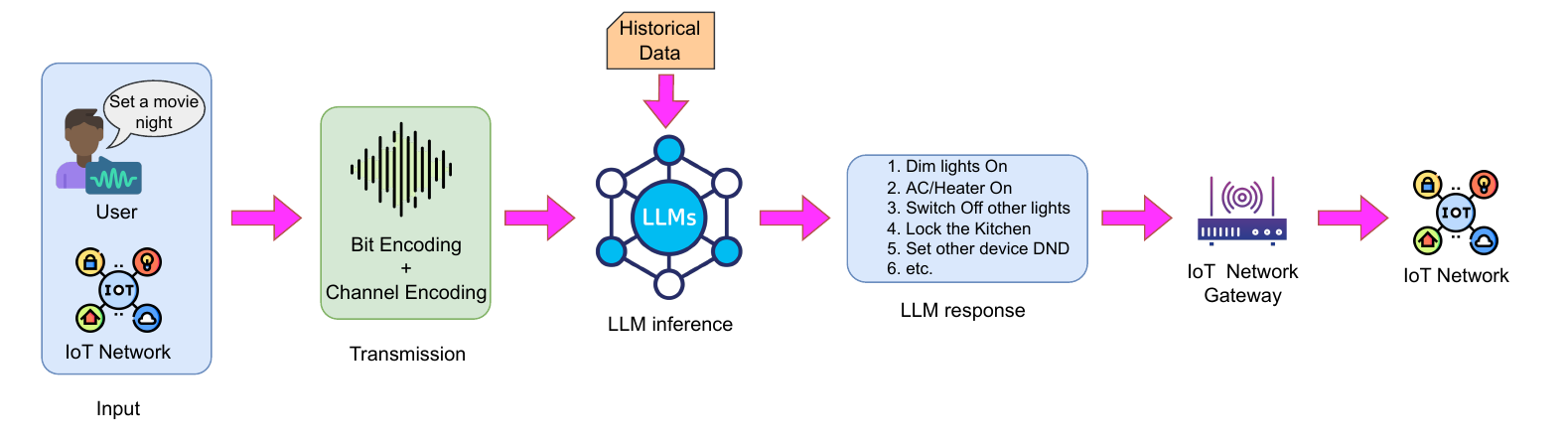}
\caption{A framework enabling LLM-based semantic communication in Edge-based IoT system}
\label{fig_2}
\end{figure*}

%As the devices used in IoT networks are resource-constraints in terms of processing capacity, memory, and power, therefore, running LLMs on devices is not feasible. As a solution to this problem, we can consider that the border router (BR) or the gateway is connected to the Edge computing devices with enough resources to infer user requests using LLMs. In brief, LLMs are deployed at the edge of the networks, and an edge device is connected to the gateway or BR of the IoT networks. The input can be given to LLMs in terms of text, and voice. However, as shown in Figure~\ref{fig_2}, along with input given by the user, LLMs can also be provided with the historical data. For example, along with the voice command to \textit{setup a movie night}, previous historical data from the same user can be considered to adjust the temperature. %We can also consider another example instead of sending a very high resolution image from the door camera, we can send the edges of the picture taken by the camera along with some textual information, such as \textit{a fair-colour man wearing a blue t-shirt with a tensed face}. This approach requires comparatively much less bandwidth to transmit a low-resolution picture and textual information than sending a very high-resolution picture. Similarly, using the previous example, the user needs to send a single command instead of six different commands to do the same activity. Based on the input provided by the user, LLMs generate commands for the actuators of the IoT network that is connected to the edge device via IoT gateway or BR. 
IoT devices are often resource-constrained regarding processing capacity, memory, and power. Consequently, running LLMs directly on these devices is not technically feasible. To address this challenge, we propose a framework where an IoT network's border router (BR) or gateway is connected to an Edge computing device equipped with sufficient resources to infer user requests using LLMs. Note that, the gateway of an IoT network is the device that connects IoT networks with different protocols such as ZigBee, Bluetooth, or 6TiSCH to the IPv6-based traditional Internet connectivity such as Ethernet, WiFi, or 5G, facilitating basic data processing and decision-making, protocol translation, and secure communication between IoT devices and the Internet.

In our proposed framework, LLMs are deployed at the edge of the network as part of semantic communication, and an edge device is connected to the gateway or BR of the IoT network. IoT devices attached to sensors collect information from the environment and transmit their data to the gateway using multi-hop (mesh topology) or single-hop (star topology) communication networks. The gateway then forwards the data to the edge computing device for processing by the LLM deployed at the edge. Note that some edge devices can also serve as IoT gateways. Apart from the input provided by the IoT devices, users can also provide their input directly to the LLMs in the form of text, video, or voice commands. For example, in smart home environments, users can send voice commands to perform different activities. The transmission of data from the IoT devices to the gateway or edge follows traditional source and channel encoding. Our framework assumes that feature restoration of semantic communication is done at the receiver end, \textit{i.e.,} the edge.

In addition to sensory input and user input, the LLM model is fed with historical data, as illustrated in Figure~\ref{fig_2}, for better inference. For instance, when a user issues a voice command to set up a movie night, our proposed framework can consider previous historical data from the user to adjust the temperature. Thus, historical data can be used to produce better inferences by the LLM. Apart from improving the performance of the LLM model using historical data, we can increase the bandwidth utilization of the overall IoT network. For this, we can consider a scenario where instead of sending a high-resolution image from the door camera, we send only the edges of the picture taken by the camera along with some textual information, such as ``\textit{a fair-colored man wearing a blue t-shirt with a tense face}'' \cite{nam2023languageorientedcommunicationsemanticcoding}. The receiver can generate a high-resolution and similar picture using the received edges-based image and text. This approach requires significantly less bandwidth to transmit a low-resolution picture and textual information compared to sending a high-resolution picture. By reducing such data load in the IoT network, these methods can conserve network resources and minimize latency, leading to faster data transmission and processing times. Additionally, it reduces the energy consumption of both the transmitting and receiving devices along with the relay (i.e., intermediate forwarding) nodes, which is crucial for maintaining the battery life of IoT devices and the overall sustainability of IoT networks. Similarly, as mentioned before, users need to send a single command instead of multiple different commands to perform the same activities in LLM-based IoT networks. Based on the input provided by the user or input collected from the IoT network, LLMs generate commands for the actuators of the IoT network. This streamlined communication improves system efficiency, enhances user experience by simplifying interactions, and reduces the potential for command errors. Furthermore, the ability to generate and execute complex commands through natural language processing enhances the scalability of IoT systems, making them more adaptable to varied applications and user needs.

\begin{figure*}[!t]
\centering
\includegraphics[width=.85\textwidth]{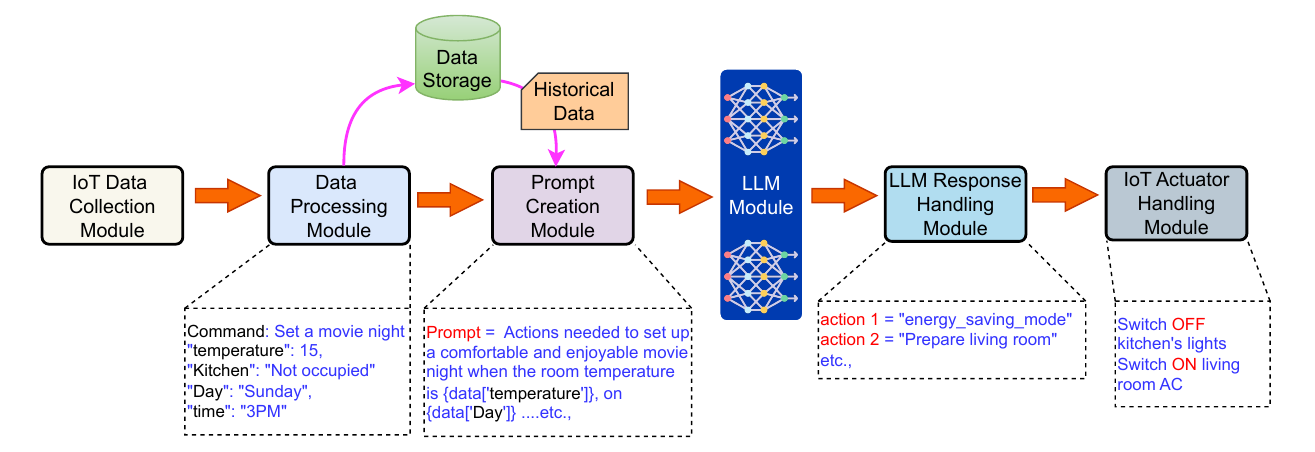}
\caption{Proposed framework is divided into multiple structured modules that handle data collection, processing, prompt creation, response handling, and actuator management.}
\label{fig_3}
\end{figure*}

To simplify our proposed framework to facilitate efficient and intelligent interactions within IoT ecosystems, we divide it into multiple structured modules that handle data collection, processing, prompt creation, response handling, and actuator management. These modules are shown in Figure~\ref{fig_3} and discussed as follows,

\textbf{IoT Data Collection Module}: This module is responsible for gathering data from various IoT devices including the command received from the users. This module ensures real-time data collection from IoT devices and users.

\textbf{Data Processing Module}: Once the data is collected, it is processed to extract relevant information and convert it into a suitable format for further analysis. This module applies filtering, aggregation, and transformation techniques to ensure the data is ready for the next stages.

\textbf{Storage Module}: This module is responsible for maintaining historical data which can be fed to the LLM for referencing past information, providing context and improving the accuracy of semantic communication.

\textbf{Prompt Creation Module}: This module generates prompts based on the processed data from the data processing module and uses historical information stored in the Storage Module. These prompts are generated to facilitate meaningful queries and interactions with the LLM.

\textbf{LLM Module}: The core component of the framework, the LLM Module, utilizes advanced language models to interpret prompts and generate coherent, context-aware responses. This module leverages the capabilities of LLMs to understand and process natural language queries effectively.

\textbf{LLM Response Handling Module}: This module processes the responses generated by the LLM. It performs further processing to refine and contextualize the output so that it can easily feed to the next module.

\textbf{IoT Actuator Handling Module}: The final module in the framework, the IoT Actuator Handling Module, translates the processed responses into actionable commands for IoT devices. It ensures that the system's outputs are effectively communicated within the IoT network, enabling intelligent and automated control.

Note that these modules can be easily integrated to work with current IoT infrastructures, enhancing their capabilities without designing the entire IoT system from scratch. For example, an LLM can interface with existing smart home assistants and IoT devices, using APIs and standard communication protocols to understand and execute user commands. 

\subsection{Advantages of using LLM for Semantic Communication in IoT Networks}
\textbf{Efficient Data Transmission}: Semantic communication focuses on transmitting the meaning or the semantics of the data rather than the raw data itself. This can significantly reduce the amount of data that needs to be transmitted, leading to more efficient use of bandwidth and lower energy consumption of the senders, which is crucial for many IoT devices that operate on limited power sources.

% \textbf{Improved Interoperability}: IoT ecosystems often involve a wide range of devices and systems from different manufacturers, which may use different protocols and data formats. semantic communication can help bridge these differences by providing a common understanding of the data, thus enhancing interoperability among heterogeneous devices.

\textbf{Enhanced Decision Making}:  Semantic communication can improve the quality of decision-making processes in various IoT applications, such as smart cities, industrial automation, and healthcare by focusing on the meaning of the data, which can lead to more relevant and actionable information. Additionally, semantic communication enables IoT systems to understand the context in which data is generated and used. This context-awareness can lead to the development of more intelligent and adaptive services that can respond dynamically to changing conditions.

\textbf{Quicker responsiveness}: As LLM in semantic communication can extract information in depth from minimal input size, therefore, it helps in quicker decision-making as transmission timing, processing timing, and queuing timing get reduced with the size of the data that has to be transmitted.  

%In scenarios where quick decision-making is essential, such as in autonomous vehicles or industrial control systems, semantic communication can reduce latency by minimizing the amount of data processing and transmission required. 

% \textbf{Enhanced Security and Privacy}: By understanding the semantics of the data, IoT systems can apply more sophisticated security and privacy measures, ensuring that sensitive information is protected while still enabling the necessary data exchanges for efficient operation.

\textbf{Enhanced User Experience}
Users can issue simple, natural language commands, which the system interprets and executes accurately, thereby minimizing the likelihood of errors. By considering historical data, the system can further consider user preferences and adjust settings accordingly, leading to a more personalized and intuitive experience. Thus, natural language interactions powered by LLMs make it easier for users to communicate with IoT devices, enhancing the overall user experience.

\section{Use Cases}
This section discusses some of the examples of how LLMs can be integrated into Edge-based IoT systems as a part of semantic communication:

\textbf{Smart Home Assistants}: 
LLMs can make smart home assistants much more powerful by enabling them to understand and respond to more natural and context-aware interactions. For instance, with the help of an LLM, a smart assistant can interpret and execute complex voice commands. If a user says, ``\texttt{Turn off the lights in the living room after dinner},'' the LLM can understand the context and timing, and then control the lights accordingly. This means users can speak more naturally and still have their commands accurately followed, making the interaction with smart home devices smoother and more intuitive. LLMs can also manage multiple IoT devices, like lights, thermostats, and security systems, seamlessly integrating their functions based on user preferences and context. Furthermore, using historical data enables the system to learn user preferences and habits over time. For example, it can adjust temperature settings based on past behaviour or suggest frequently used commands. Additionally, historical data can be used to understand the context of the command in a better way. For instance, if it knows that a user typically dims the lights and plays soft music in the evening, it can proactively make these adjustments without explicit commands. Overall the entire system can create a more personalized and intuitive user experience.

\textbf{Industrial IoT}:
In industrial IoT, IoT devices are deployed to continuously monitor machinery and equipment. Sensor-attached IoT devices collect vast amounts of data, including temperature readings, vibration levels, and operational cycles, etc. By understanding the intricate patterns and contexts within the data, LLMs can identify possible issues that may not be determined by the existing system and accordingly, LLMs can suggest detailed maintenance schedules and recommendations.  Such preemptive replacement recommendations before the actual failure and real-time alerts can significantly improve the efficiency of the Industrial ecosystems by preventing unexpected downtime. The efficiency of industries can be further improved by allowing LLMs to optimize the overall maintenance strategy by learning from historical data. They can identify trends such as recurring failures or seasonal variations in equipment performance, allowing for more strategic planning. Furthermore, LLM-generated responses can simplify complex information, making it easy for non-technical individuals to understand and act on maintenance needs. For example, instead of presenting raw data, the LLM can provide clear, straightforward instructions like, ``\texttt{Machine A1025 needs a check-up due to unusual vibrations. Schedule a maintenance session this week.}'' This helps non-technical staff also to understand the situation and take appropriate action.

\textbf{Smart Healthcare}:
LLMs can be seamlessly integrated with IoT devices in healthcare to enhance real-time patient monitoring and support. For instance, wearable sensors can continuously gather vital patient data such as heart rate, body temperature, and activity levels. An LLM can then analyze this data to detect any irregularities, predict potential health issues, and provide tailored health advice. If a patient’s heart rate becomes irregular, the system can immediately alert healthcare providers, offering insights into possible causes and suggesting appropriate actions. This proactive approach ensures timely medical intervention, potentially preventing more serious health complications.

Moreover, LLM-based semantic communication can significantly improve coordination and efficiency in medical environments such as surgical operations. For example, in a scenario where a doctor performs a complex surgery and uses voice commands to interact with various IoT-enabled medical devices and communicate with the surgical team. For example, the doctor might say, ``\texttt{Increase the oxygen level by 10\%}'' and the LLM would interpret this command, adjust the settings on the relevant device, and \textit{confirm} the change. This allows the doctor to focus on the surgery without manually adjusting equipment.

These examples illustrate how LLM-based semantic communication can add significant value to IoT systems by providing advanced data analysis, context-aware interactions, and predictive capabilities.

\section{Open Challenges and Issues}
Even though LLMs under the umbrella of semantic communication in Edge-based IoT systems can revolutionize the entire communication domain, there are a variety of challenges to be tackled before they can be used in real applications. This section discusses the two major open issues for future investigations.

\subsection{Privacy and Security}
One of the significant challenges when integrating LLMs with semantic communication, edge computing, and IoT systems is ensuring their dependability and reliability. One major concern is safeguarding these models against malicious attacks, as attacks can lead to unexpected behaviours and unreliable outputs. Moreover, privacy issues are prominent, as LLMs can be susceptible to revealing sensitive information embedded within the training data. Furthermore, processing sensitive data locally on edge devices can help enhance privacy and security, but it also poses challenges. Protecting data from breaches is tough, especially in large-scale deployments across numerous edge devices. Robust encryption, secure data handling practices, and regular security updates are essential but can be resource-intensive. Hence, ensuring these security measures is vital for the secure and reliable operation of semantic communication systems that leverage LLMs in edge computing-based and IoT networks. Addressing these challenges requires a combination of advanced technical solutions, robust security measures, and ongoing optimization efforts to make the integration of LLMs, semantic communication, and edge-based IoT systems both practical and effective.

\subsection{Data Heterogeneity and Quality}
%IoT systems generate vast amounts of data from diverse sources, often with varying levels of quality and consistency. Ensuring that LLMs can effectively handle and interpret heterogeneous data is challenging. Poor-quality or inconsistent data can lead to inaccurate model outputs. Developing robust preprocessing techniques and data integration methods is essential to maintain high-quality, reliable model performance.
IoT systems generate vast and diverse datasets from numerous interconnected devices, each producing data in various formats and frequencies, which presents significant challenges for LLMs. Data heterogeneity involves integrating structured data from databases, unstructured data from text or logs, time-series data from sensors, and multimedia data like images, requiring techniques such as multi-modal learning, data fusion, and semantic mapping to unify and interpret these varied inputs effectively. Meanwhile, ensuring data quality is crucial, as issues like noise, missing values, and inconsistencies can severely impact model performance. Addressing the issues with data quality involves pre-processing methods such as data cleaning, normalization, feature engineering, and anomaly detection, alongside robust integration strategies like data aggregation and continuous monitoring. These approaches are essential for maintaining high-quality, reliable LLM outputs out of unstructured and heterogeneous IoT data.

\section{Conclusion}

In this article, we discussed how semantic communication can leverage LLMs to enhance the efficiency of edge-based IoT systems. We proposed a framework for designing such IoT systems and briefly covered each module involved in the proposed framework. We also examined the benefits of using LLMs for semantic communication in IoT networks, along with several practical use cases. Finally, we addressed some open challenges that need to be resolved before implementing such IoT systems.

\bibliographystyle{IEEEtran}

% Include the .bbl file directly

 % Assuming your .bbl file is named references.bbl
\end{document}